\documentclass[twocolumn,prl,showpacs]{revtex4}

\usepackage{graphicx}
\usepackage{dcolumn}
\usepackage{float}
\usepackage{amsmath}

\newcommand{\comment}[1]{}

\begin{document}

\title{Evidence for a pseudogap in underdoped Bi$_{2}$Sr$_{2}$CaCu$_{2}$O$_{8+\delta}$
and YBa$_{2}$Cu$_{3}$O$_{6.50}$ from in-plane optical conductivity
measurements}

\author{J. Hwang$^{1}$}
\email{jhwang@phys.ufl.edu}
\author{J. P. Carbotte$^{1,2}$}
\author{T. Timusk$^{1,2}$}

\affiliation{$^{1}$Department of Physics and Astronomy, McMaster
University, Hamilton, Ontario L8S 4M1, Canada}
\affiliation{$^{2}$The Canadian Institute for Advanced Research,
Toronto, Ontario M5G 1Z8, Canada}

\date{\today}

\begin{abstract}
The real part of the in-plane optical self-energy data in
underdoped Bi$_{2}$Sr$_{2}$CaCu$_{2}$O$_{8+\delta}$ (Bi-2212) and
ortho II YBa$_{2}$Cu$_{3}$O$_{6.5}$ contains new and important
information on the pseudogap. Using a theoretical model approach
we find that the density of state lost below the pseudogap
$\Delta_{pg}$ is accompanied with a pileup just above this energy.
The pileup along with a sharp mode in the bosonic spectral
function leads to an unusually rapid increase in the optical
scattering rate and a characteristically sloped peak in the real
part of the optical self-energy. These features are not found in
optimally doped and overdoped samples and represent a clearest
signature so far of the opening of a pseudogap in the in-plane
optical conductivity.
\end{abstract}

\pacs{74.25.Gz, 74.62.Dh, 74.72.Hs}

\maketitle


A striking feature of the underdoped cuprates, not present in
conventional metals, is the appearance of a
pseudogap\cite{timusk99} below a characteristic temperature $T^*$.
In general, the opening of a pseudogap $\Delta_{pg}$ leads to a
reduction in the electronic density of states (DOS) around the
Fermi energy. Since, as yet, there is no full understanding of the
exact origin of the pseudogap in the cuprates, an important issue
is how it should be modelled, its width in frequency, its behavior
as a function of doping and temperature and what happens to the
spectral weight  in the gap. While the first evidence of a
pseudogap in the cuprates came from NMR
experiments\cite{warren89,alloul89} that showed a gap in the spin
excitation spectrum, a gap in the charge excitations was found in
the c-axis conductivity of underdoped YBCO by Homes {\it et
al.}\cite{homes93}. The conductivity in this direction is
incoherent, and like c-axis tunneling, is expected to be
proportional to the density of states. Homes {\it et al.}
concluded that the spectral weight lost in the pseudogap region
was recovered at much higher frequencies. A similar conclusion was
reached by Loram {\it et al.} from specific heat measurements
\cite{loram98}. In contrast, Yu {\it et al.} from a recent study
of c-axis conductivity of underdoped RBa$_{2}$Cu$_{3}$O$_{6.5}$
(R=Y, Nd, and La) extending the measurements to higher frequency
found that spectral weight was recovered~\cite{Yu07}. Also
tunneling data suggests that the "coherence peak" that signals a
tunneling conductance that is enhanced over the background
persists well into the normal state in Bi-2212 \cite{renner98}
suggesting a recovery of gapped states.

The pseudogap is more difficult to identify in the ab-plane
transport. While it was clear very early on that the ab-plane
absorption showed an unmistakable decrease at energies below
$\Delta_{pg}$\cite{orenstein90,rotter91,basov96,puchkov96,hwang06},
a clear interpretation in terms of a pseudogap was problematic
since it is difficult to separate the effects of a pseudogap (a
pseudogap in the density of states) and a gap in the spectrum of
inelastic excitations (a spin gap) since both reduce absorption at
low frequencies. Further, studies of the real part of the
conductivity did not show the expected reduction in optical
spectral weight below the pseudogap energy\cite{santander02}. This
remains unexplained. In this letter we find that by focussing not
on features in the optical conductivity, but on the real and
imaginary parts of the optical self-energy we {\it are able to}
separate the contributions of the pseudogap and inelastic
scattering. Furthermore, simulations with simple models show that
spectral features in the density of states are less spread out in
the self-energy than in the conductivity making small changes
easier to estimate accurately. We can trace to missing states
below the gap to the region just above it.

It has become common in the optical literature to analyze
reflectivity measurements in terms of the real and imaginary parts
of the optical conductivity
$\sigma(\omega)=\sigma_1(\omega)+i\sigma_2(\omega)$ and present
the results for the optical scattering rate $1/\tau^{op}(\omega)$
as a function of $\omega$. The extended Drude formula, which is
valid for correlated electrons, can be written as $ 4 \pi
\sigma(\omega)\equiv i\omega_{p}^2
/[\omega-2\Sigma^{op}(\omega)]$, where $\omega_p$ is the plasma
frequency and
$\Sigma^{op}(\omega)\equiv\Sigma_1^{op}(\omega)+i\Sigma_2^{op}(\omega)$
is by definition the optical self-energy. Its imaginary part gives
the optical scattering rate and its real part is related to the
optical mass re-normalization.  The real and imaginary parts of
$\Sigma^{op}(\omega)$ are related by Kramers-Kronig (K-K)
transformations. They play, for the optical conductivity, a role
similar to that played by the quasiparticle self-energy
$\Sigma^{qp}(\omega)$ in angle-resolved photoemission spectroscopy
but these are different quantities although they are closely
related~\cite{carbotte05,schachinger03}.

The real part of $\Sigma^{op}(\omega)$ obtained from the in-plane
conductivity in the highly underdoped
regime\cite{hwang06,hwang04,hwang07} shows a particularly clear
and striking signature of the pseudogap modelled just as was done
from the specific heat\cite{loram98} by a simple reduction in
electronic density of states in the vicinity of the Fermi energy.
Contrary to the specific heat case, however, we find evidence that
these missing states in DOS pile up in the region just above the
pseudogap energy. While the exact shape taken by the quasiparticle
DOS around the Fermi energy will in principle be reflected in the
shape of the real part of the optical self-energy, these details
are of secondary importance.

\begin{figure}[t]
  \vspace*{-1.0 cm}%
  \centerline{\includegraphics[width=3.5 in]{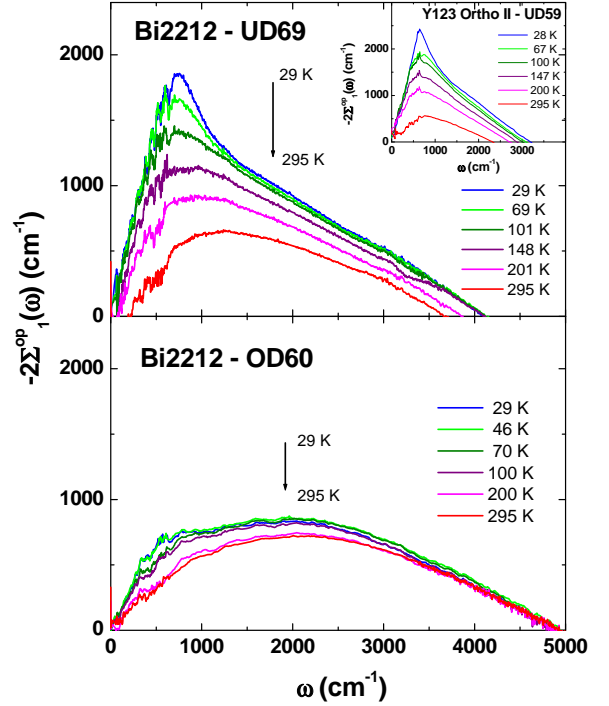}}%
  \vspace*{-1.0 cm}%
\caption{(color online) Minus twice the real part of the optical
self-energy $-2\Sigma^{op}(\omega)$ vs. $\omega$ for underdoped
(top frame) and overdoped Bi-2212 (bottom
frame)\cite{hwang04,hwang07}. The inset shows the corresponding
results for underdoped ortho II YBCO\cite{hwang06}.}
 \label{Fig1}
\end{figure}

In Fig.~\ref{Fig1} we compare results for
$-2\Sigma^{op}_1(\omega)$ vs. $\omega$ obtained in ref.
\cite{hwang04,hwang07} for an underdoped Bi-2212 sample (top
frame) with an overdoped sample with approximately the same
critical temperature (bottom frame). The evolution of the curves
seen as the temperature is lowered is quite different in the two
cases. The overdoped samples show only small changes while the
underdoped one shows the growth of a sharp triangle-like peak
around 750 cm$^{-1}$. We will argue that this sharp peak is a
result of a pseudogap opening with a recovery region just above it
while at the same time a sharp resonance exists in the
electron-boson spectral density $I^2\chi(\omega)$ which we believe
to be due to spin fluctuations and not phonons. The inset in the
top frame of Fig. 1 shows results from the highly underdoped ortho
II compound YBa$_{2}$Cu$_{3}$O$_{6.5}$\cite{hwang06} in which
every second chain is full and the others are empty, leading to a
perfect stoichiometric sample. The same characteristic temperature
evolution is seen in YBCO as in Bi-2212. This suggests that this
sharp peak is a universal feature of the underdoped cuprates.

To establish our central point it will be sufficient to use
approximate, but surprisingly accurate, formulas that relate the
real and imaginary parts of the scattering rates to the
electron-boson spectral density $I^2\chi(\omega)$ for zero
temperature from the work of Mitrovic {\it et
al.}\cite{mitrovic85} and
others\cite{allen71,shulga91,sharapov05,knigavko05}. At $T=0$ we
have\cite{mitrovic85},
\begin{equation}
\begin{split}
-2\Sigma_1^{op}(\omega)\! \cong \! \frac{2}{\omega}
\!\int^{\infty}_0 \!\!d\Omega I^2\chi(\Omega) \!\int^{\infty}_0
\!\!d\omega'\tilde{
N}(\omega')\\
\times
\ln\Big{|}\frac{(\omega'+\Omega)^2}{(\omega'+\Omega)^2-\omega^2}\Big{|}\nonumber
\end{split}
\end{equation}
\begin{equation} -2\Sigma_2^{op}(\omega)\!\equiv\!
\frac{1}{\tau^{op}(\omega)}\!\cong \!
\frac{2\pi}{\omega}\!\!\int^{\omega}_0 \!\!\!\!d\Omega
I^2\chi(\Omega)\!\!\int^{\omega-\Omega}_0
\!\!\!\!\!\!\!d\omega'\tilde{N}(\omega')\nonumber
\end{equation}
where $\tilde{N}(\omega)$ is the renormalized electronic density
of states symmetrized about the Fermi energy. It contains the
pseudogap and, for finite bands, would decay rapidly at the
renormalized band edge. There is a cutoff frequency, $\Omega_c$ on
the spectral density. Here we used $\Omega_c$ = 4000 cm$^{-1}$ .
%
%

We begin with the following observation. For a constant density of
states and coupling to a single mode of energy $\omega_E$ {\it
i.e.} $I^2\chi(\omega)=A_0\delta(\omega-\omega_E)$ with $A_0$
setting the scale, the optical scattering rate
($1/\tau^{op}(\omega)$) is zero for $\omega < \omega_E$ and equal
to $2 \pi A_0[(\omega-\omega_E)/\omega]$ for $\omega \geq
\omega_E$ which starts from zero at $\omega=\omega_E$ and
saturates to $2 \pi A_0$ only for $\omega \gg \omega_E$. If, in
addition, there is a full gap $\Delta_{pg}$ (the pseudogap) in the
symmetrized density of electronic states $\tilde{N}(\omega)$,
$1/\tau^{op}(\omega)$ starts instead from zero at
$\omega=\omega_E+\Delta_{pg}$ and rise towards saturation
according to the modulating factor
$(\omega-\omega_E-\Delta_{pg})/\omega$ which effectively rises
more slowly with $\omega$ than does $(\omega-\omega_E)/\omega$.
This is seen in the middle frame of Fig. 2 where we show
$-2\Sigma^{op}_2(\omega)$ for three cases, A, B, C all based on
the electron-boson spectral density  shown in the inset,
$I^2\chi(\omega)=A_p/[\sqrt{2\pi}(d/2.35)]
e^{-(\omega-\omega_{p})^2/[2(d/2.35)^2]} +A_s\:
\omega/[\omega_{sf}^2+\omega^2]$, where $A_p=$ 200, $\omega_{p}=$
250 cm$^{-1}$, $d=$ 10 cm$^{-1}$, $A_s=$ 100 cm$^{-1}$, and
$\omega_{sf}=$ 500 cm$^{-1}$. It has a peak at $\omega=250$
cm$^{-1}$ and a broad background extending to 4000 cm$^{-1}$,
modelled on results for underdoped ortho II YBCO\cite{hwang06}.
The curve A includes no pseudogap while curve B has $\Delta_{pg} =
550$ cm$^{-1}$ and starts from zero at 800 cm$^{-1}$ in contrast
to curve A which starts at 250 cm$^{-1}$. Comparing with curve A
shows that a pseudogap leads to an optical scattering rate which
rises less steeply from zero than it would with $\Delta_{pg}=0$. A
slower rise would also arise if instead of using a delta function
at $\omega_E$ for $I^2\chi(\omega)$ part of the weight (not shown
here) was distributed to a region of higher frequencies. The two
curves for $-2\Sigma^{op}_2(\omega)$ as well as C continue to rise
at the largest photon energies shown because we have included a
background in $I^2\chi(\omega)$ which extends to 4000 cm$^{-1}$.
The curve C includes a full pseudogap just as in curve B but now
it is also assumed that the lost states in $\tilde{N}(\omega)$
reappear just above $\Delta_{pg}$ in the interval $\Delta_{pg}$ to
$2\Delta_{pg}$. These extra electronic states lead to increased
scattering  and the curve C rises from zero much more rapidly than
curve B. Also at the energy marking the end of the enhanced
density of states (recovery region) there is a kink in
$1/\tau^{op}(\omega)$ beyond which the rate of increase in
$1/\tau^{op}(\omega)$ becomes less rapid.
%
%
\begin{figure}[t]
  \vspace*{-1.0 cm}%
  \centerline{\includegraphics[width=3.5 in]{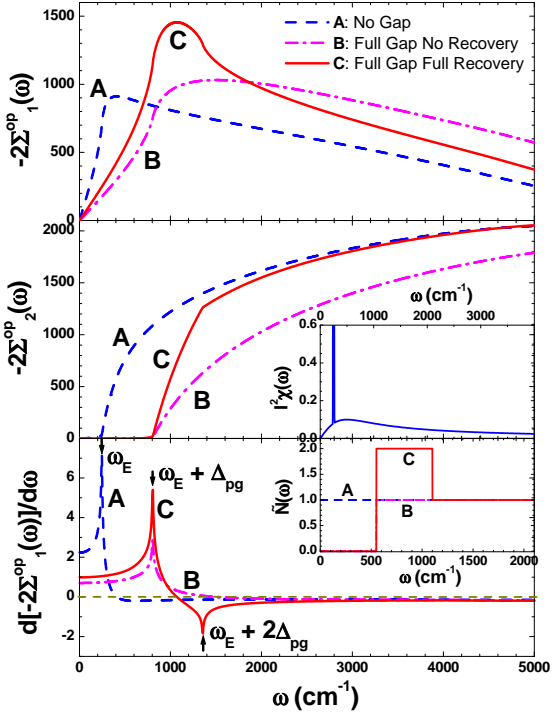}}%
  \vspace*{-1.0 cm}%
\caption{(color online) Model results for
$-2\Sigma^{op}_1(\omega)$ vs. $\omega$ (top frame) and
$-2\Sigma^{op}_2(\omega)$ vs. $\omega$ (middle frame). The
electrons are coupled to bosons with electron-boson spectral
weight shown in the inset of the middle frame. A, B and C curves
are with no pseudogap, pseudogap without recovery region, and
pseudogap with recovery, respectively. The corresponding density
of states, $\tilde{N}(\omega)$, is shown in the inset (bottom
frame) with the same line types. In the bottom frame we show the
derivative, $d[-2\Sigma^{op}_1(\omega)]/d\omega$ vs. $\omega$.}
  \label{Fig2}
\end{figure}
%
%
\begin{figure}[t]
  \vspace*{-1.7 cm}%
  \centerline{\includegraphics[width=3.5 in]{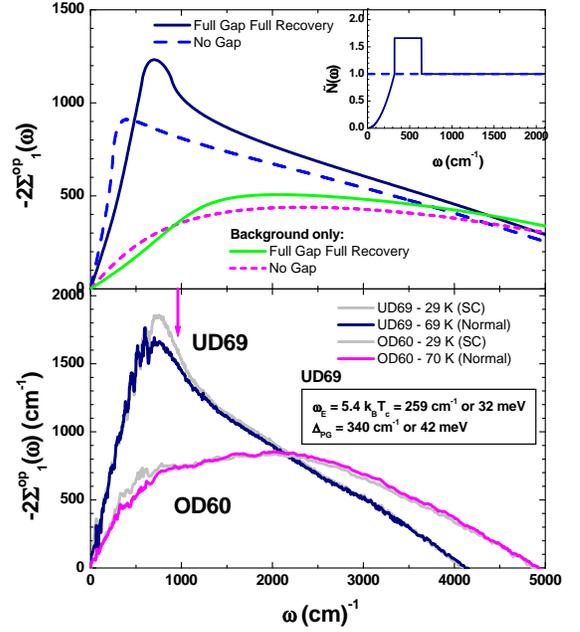}}%
  \vspace*{-1.7 cm}%
\caption{(color online) Comparison of $-2\Sigma^{op}_1(\omega)$
vs. $\omega$ for underdoped Bi-2212 and overdoped Bi-2212 samples,
bottom frame. The top frame shows results of model calculations.
Solid (dark-blue) and dashed (blue) curves are with pseudogap and
recovery and without gap (see corresponding quadratic behavior
$\tilde{N}(\omega)$ curves in the inset) for the $I^2\chi(\omega)$
shown in the inset to Fig.~\ref{Fig2}. Dashed (purple) and solid
(green) curves are obtained when the electron-boson spectral
density, just has a continuous background.}
  \label{Fig3}
\end{figure}

In the top frame of Fig.~\ref{Fig2} we show the real part of
$\Sigma^{op}(\omega)$ which corresponds to the K-K transformation
of the optical scattering rates of the middle frame. For the curve
A there is no singularity at $\omega_E$. Rather there is a very
broad maximum at $\omega \sim \sqrt{2}\omega_E$ followed by a slow
drop to about 1/3 its maximum value at 4000 cm$^{-1}$. If, on the
other hand, we include a pseudogap without a recovery region above
it, we get curve B, for which the peak falls at $\sim \sqrt
2(\omega_E + \Delta_{pg}$). Including a recovery region in the
density of states changes the shape of $\Sigma^{op}_1(\omega)$
significantly as we can see in curve C. Here we include a full gap
below $\Delta_{pg}$ and displaced states just above as shown in
the solid (red) curve of the inset on the bottom panel of
Fig.~\ref{Fig2} where our model $\tilde{N}(\omega)$ is shown. In
this cases $-2\Sigma^{op}_1(\omega)$ develops a much more
pronounced peak between $\omega_E + \Delta_{pg}$ and $\omega_E +
2\Delta_{pg}$, points marked by maximum slopes. This structure is
seen both in underdoped Bi-2212 and YBCO samples at low
temperature normal state of Fig.~\ref{Fig1} and is strong evidence
for a pileup of the spectral weight lost below $\Delta_{pg}$, in
the region $\Delta_{pg} < \omega < 2\Delta_{pg}$. The above
characteristics are a clear signature of pseudogap formation in
the in-plane conductivity.

The feature of the model curves displayed in the top frame of
Fig.~\ref{Fig2} which we have just described and which are most
important for the interpretation of experiments, shows up even
more clearly in the {\it derivative} of the real part of the
optical self-energy $d[-2\Sigma^{op}_1(\omega)]/d\omega$. Results
are shown in the bottom panel of Fig.~\ref{Fig2}. The curve A
shows maximum slope at the energy $\omega_E$ of the peak in
$I^2\chi(\omega)$ (see inset in the middle frame of
Fig.~\ref{Fig2}) and it is zero at $\sim \sqrt{2}\omega_E$. Curve
B has a maximum at $\omega_E+\Delta_{pg}$ and a zero at $\sim
\sqrt{2}(\omega_E+\Delta_{pg})$ while curve C peaks at
$\omega_E+\Delta_{pg}$ and shows a minimum at
$\omega_E+2\Delta_{pg}$ indicating the end of the recovery region
in our model of the effective electronic density of states,
$\tilde{N}(\omega)$.

In the lower panel of Fig.~\ref{Fig3} we repeat our experimental
results for the underdoped and overdoped samples of
Fig.~\ref{Fig1} but for clarity, show only the two lowest
temperatures. In the top frame we show four curves. One set is
based on the electron-boson spectral density shown in the inset of
Fig.~\ref{Fig2}. The other is obtained when only the background is

kept in $I^2\chi(\omega)$. In Bi-2212 series we have found that in
the overdoped regime there is no prominent peak in the
electron-boson spectral density, only a background
remains\cite{carbotte05,schachinger03,schachinger06,hwang06a}. For
this figure we have used quadratic pseudogap shown as the
(dark-blue) solid curve in the inset of Fig.~\ref{Fig3}. Starting
with the two curves for the background only, $I^2\chi(\omega)=
100\:\omega/[500^2+\omega^2]$, we see that including a pseudogap
does not change the curve for $\Sigma^{op}_1(\omega)$ much in
agreement with the data and there is, in this sense, no clear
signature of $\Delta_{pg}$ in such spectra. On the other hand
contrasting with the other two curves, the pseudogap now clearly
changes the shape of $\Sigma^{op}_1(\omega)$ and brings it much
closer to the experimental results for the underdoped sample
(lower panel). The peak in the (dark-blue) solid curve can be made
sharper and resemble even more a logarithmic singularity by
decreasing the width of the recovery region. This arises because
the sharp rise seen in $1/\tau^{op}(\omega)$ (curve B) of
Fig.~\ref{Fig2} middle frame, can be made even steeper and come
closer to a pure vertical rise. We conclude that the striking
difference observed in $\Sigma_1^{op}(\omega)$ between underdoped
and overdoped samples can be understood to result from the absence
of a pseudogap and a sharp peak in $I^2\chi(\omega)$ for the
overdoped case, and a combination of a peak and pseudogap in the
underdoped case. Both are needed to get agreement with the data,
including a recovery of spectral weight in the electronic density
of states in the region just above the pseudogap with the end of
the recovery region marked by a kink in $\Sigma^{op}_1(\omega)$
around 940 cm$^{-1}$ for Bi-2212 (purple arrow in lower panel of
Fig.~\ref{Fig3}) and similarly 970 cm$^{-1}$ for YBCO. If we
assume the position of the peak in $I^2\chi(\omega)$ to be set by
the neutron resonance energy given by $\Omega_{res}=5.4 k_B T_c$
\cite{he01} we conclude that the pseudogap has a value of 43 meV
in Bi-2212 and nearly the same 45 meV in YBCO and that the
recovery region is approximately $\Delta_{pg}$ wide. We note that
our suggestion that the recovery of spectral weight occurs
immediately above the gap region in the normal state just above
$T_c$, is in accord with tunneling spectra~\cite{renner98}.

In summary, the real part of the optical self-energy,
$-2\Sigma^{op}_1(\omega)$, defined through a generalized Drude
formula at low temperature in the normal state, for the underdoped
case, shows characteristic triangular behavior around 750
cm$^{-1}$ not seen in overdoped samples. This behavior is traced
to the existence of a pseudogap $\Delta_{pg}$ in the electronic
density of states with states lost below $\Delta_{pg}$ recovered
in the region immediately above it as well as the existence of a
peak in the electron-boson spectral density responsible for the
inelastic scattering. By contrast, in the overdoped samples no
pseudogap opens and the fluctuation spectrum $I^2\chi(\omega)$ has
no resonance peak but rather consists only of a reasonably flat
background. Criteria to recover a value of the pseudogap and the
end of the recovery region are given and applied to underdoped
Bi-2212 and ortho II YBCO. The signature of the pseudogap is much
clearer in the optical self-energy than it is in the in-plane
conductivity, the quantity used in previous studies.

\acknowledgments This work has been supported by the Natural
Science and Engineering Research Council of Canada (NSERC) and the
Canadian Institute for Advanced Research (CIAR). We acknowledge
the contribution of Hiroshi Eisaki and Genda Gu for the Bi-2212
crystals and Ruixing Liang, Doug Bonn, and Walter Hardy for the
YBCO crystal.

\end{document}